\documentclass[twocolumn,aps,showpacs,prl,superscriptaddress]{revtex4} 

\usepackage{sidecap}
\usepackage{ulem}
\usepackage{epsfig}
\usepackage{amsmath,amssymb,amsthm}
\usepackage{graphicx}
\usepackage{bm}
\usepackage{color}

\DeclareMathAlphabet{\mathpzc}{OT1}{pzc}{m}{it}

\begin{document}

\renewcommand{\textfraction}{0.00}


\newcommand{\vAi}{{\cal A}_{i_1\cdots i_n}} 
\newcommand{\vAim}{{\cal A}_{i_1\cdots i_{n-1}}} 
\newcommand{\vAbi}{\bar{\cal A}^{i_1\cdots i_n}}
\newcommand{\vAbim}{\bar{\cal A}^{i_1\cdots i_{n-1}}}
\newcommand{\htS}{\hat{S}} 
\newcommand{\htR}{\hat{R}}
\newcommand{\htB}{\hat{B}} 
\newcommand{\htD}{\hat{D}}
\newcommand{\htV}{\hat{V}} 
\newcommand{\cT}{{\cal T}} 
\newcommand{\cM}{{\cal M}} 
\newcommand{\cMs}{{\cal M}^*}
\newcommand{\vk}{\vec{\mathbf{k}}}
\newcommand{\bk}{\bm{k}}
\newcommand{\kt}{\bm{k}_\perp}
\newcommand{\kp}{k_\perp}
\newcommand{\km}{k_\mathrm{max}}
\newcommand{\vl}{\vec{\mathbf{l}}}
\newcommand{\bl}{\bm{l}}
\newcommand{\bK}{\bm{K}} 
\newcommand{\bb}{\bm{b}} 
\newcommand{\qm}{q_\mathrm{max}}
\newcommand{\vp}{\vec{\mathbf{p}}}
\newcommand{\bp}{\bm{p}} 
\newcommand{\vq}{\vec{\mathbf{q}}}
\newcommand{\bq}{\bm{q}} 
\newcommand{\qt}{\bm{q}_\perp}
\newcommand{\qp}{q_\perp}
\newcommand{\bQ}{\bm{Q}}
\newcommand{\vx}{\vec{\mathbf{x}}}
\newcommand{\bx}{\bm{x}}
\newcommand{\tr}{{{\rm Tr\,}}} 
\newcommand{\bc}{\textcolor{blue}}

\newcommand{\beq}{\begin{equation}}
\newcommand{\eeq}[1]{\label{#1} \end{equation}} 
\newcommand{\ee}{\end{equation}}
\newcommand{\bea}{\begin{eqnarray}} 
\newcommand{\eea}{\end{eqnarray}}
\newcommand{\beqar}{\begin{eqnarray}} 
\newcommand{\eeqar}[1]{\label{#1}\end{eqnarray}}
 
\newcommand{\half}{{\textstyle\frac{1}{2}}} 
\newcommand{\ben}{\begin{enumerate}} 
\newcommand{\een}{\end{enumerate}}
\newcommand{\bit}{\begin{itemize}} 
\newcommand{\eit}{\end{itemize}}
\newcommand{\ec}{\end{center}}
\newcommand{\bra}[1]{\langle {#1}|}
\newcommand{\ket}[1]{|{#1}\rangle}
\newcommand{\norm}[2]{\langle{#1}|{#2}\rangle}
\newcommand{\brac}[3]{\langle{#1}|{#2}|{#3}\rangle} 
\newcommand{\hilb}{{\cal H}} 
\newcommand{\pleft}{\stackrel{\leftarrow}{\partial}}
\newcommand{\pright}{\stackrel{\rightarrow}{\partial}}
\title{Jet suppression of pions and single electrons at Au+Au collisions 
at RHIC}

\date{\today}
 
\author{Magdalena Djordjevic}
\affiliation{Institute of Physics Belgrade, University of Belgrade, Serbia}

\begin{abstract} 
Jet suppression is considered to be a powerful tool to study the properties 
of a QCD medium created in ultra-relativistic heavy ion collisions. However, 
theoretical predictions obtained by using jet energy loss in {\it static} QCD
medium show disagreement with experimental data, which is known as the heavy 
flavor puzzle at RHIC. We calculate the suppression patterns of pions and 
single electrons for Au+Au collisions at RHIC by including the energy loss in 
a finite size {\it dynamical} QCD medium, with finite magnetic mass effects 
taken into account. In contrast to the static case, we here report a good 
agreement with the experimental results, where this agreement is robust with 
respect to magnetic mass values. Therefore, the inclusion of dynamical QCD 
medium effects provides a reasonable explanation of the heavy flavor puzzle 
at RHIC.
\end{abstract}


\maketitle 

\section{Introduction} 

Jet suppression~\cite{Bjorken} measurements at RHIC and LHC, and their 
comparison with theoretical predictions, provide a powerful tool for mapping 
the properties of a QCD medium created in ultra-relativistic heavy ion 
collisions~\cite{Brambilla,Gyulassy,DBLecture}. However, jet suppression 
predictions, done under assumption of static QCD medium~\cite{DGVW,RGH,ADSW}, 
showed a disagreement with the available data from RHIC 
experiments~\cite{PHENIX_pi0,PHENIX_e,STAR_pi0,STAR_e}. This disagreement has 
been named ``heavy flavor puzzle at RHIC''~\cite{Gyulassy_viewpoint,SE_puzzle}, 
and raised important questions about the ability of the available theories to 
model the matter created at ultra-relativistic heavy ion collisions at RHIC.

Since the suppression results from the energy loss of high energy partons 
moving through the plasma~\cite{suppression,BDMS,BSZ,KW:2004}, accurate 
computations of jet energy loss mechanisms are essential for the reliable 
predictions of jet suppression. In~\cite{MD_PRC,DH_PRL}, we developed a 
theoretical formalism for the calculation of radiative energy loss in 
realistic finite size {\it dynamical} QCD medium (see also a 
viewpoint~\cite{Gyulassy_viewpoint}), which abolished a static approximation 
used in previous models~\cite{GLV,Gyulassy_Wang,Wiedemann,WW,DG_Ind,ASW,MD_TR}. 
Furthermore, in~\cite{MD_MagnMass}, we extended the study from~\cite{MD_PRC} 
to include a possibility for existence of finite magnetic mass; this 
generalization was motivated by various non-perturbative 
approaches~\cite{Maezawa,Nakamura,Hart,Bak}, which report non-zero magnetic 
mass. These studies, together with the previously developed 
collisional energy loss formalism in finite size dynamical QCD 
medium~\cite{MD_Coll}, enable us to provide the most reliable computations of 
the energy loss in QGP so far.

In this paper, we integrate the developed energy loss formalism into a 
computational framework that can generate reliable predictions for RHIC and 
LHC experimental data. The numerical procedure includes: {\it i)} both 
collisional and radiative energy loss from the newly developed (dynamical QCD 
medium) formalism~\cite{MD_PRC,DH_PRL,MD_MagnMass,MD_Coll}, {\it ii)} 
multi-gluon fluctuations, i.e. the fact that energy loss is a 
distribution~\cite{GLV_suppress}, and {\it iii)} path length fluctuations, i.e. 
the fact that particles travel different paths in the medium~\cite{WHDG}. 
We use this framework to generate suppression predictions for pions and single 
electrons at most central 200 GeV Au+Au collisions at RHIC. The generated 
predictions are directly compared with RHIC experimental 
data~\cite{PHENIX_pi0,PHENIX_e,STAR_pi0,STAR_e}, in order to test our 
understanding of QGP created at these collisions.
\section{Computational framework}

The quenched spectra of partons, hadrons, and leptons are calculated
as in~\cite{Djordjevic:2005db,WHDG} from the generic pQCD convolution
\begin{eqnarray}
\frac{E_f d^3\sigma(e)}{dp_f^3} &=& \frac{E_i d^3\sigma(Q)}{dp^3_i}
 \otimes
{P(E_i \rightarrow E_f )}\nonumber \\
&\otimes& D(Q \to H_Q) \otimes f(H_Q \to e), \; 
\label{schem} \end{eqnarray}
where $Q$ denotes quarks and gluons. For charm and bottom, the initial quark 
spectrum, $E_i d^3\sigma(Q)/dp_i^3$, is computed at next-to-leading order
using the code from~\cite{Cacciari:2005rk,MNR}; for gluons and light quarks, 
the initial distributions are computed at leading order as 
in~\cite{Vitev:2002pf}. $P(E_i \rightarrow E_f )$ is the energy loss 
probability, $D(Q \to H_Q)$ is the fragmentation function of quark or gluon 
$Q$ to hadron $H_Q$. The last step ($f(H_Q \to e)$) is only applicable for 
heavy quarks, and it represents the decay function of hadron $H_Q$ into the 
observed single electron. We use the same mass and factorization scales as
in~\cite{Vogt} and employ the CTEQ5M parton densities~\cite{Lai:1999wy}
with no intrinsic $k_T$. As in~\cite{Vogt} we neglect shadowing of
the nuclear parton distribution.

We assume that the final quenched energy $E_f$ is large enough that 
the Eikonal approximation can be employed. We also assume that in Au+Au 
collisions, the jet to hadron fragmentation functions are the same as in 
$e^+e^-$ collisions. This assumption is expected to be valid in the deconfined 
medium case, where hadronization of $Q\rightarrow H_Q$ cannot occur 
until the quark emerges from the QGP. 

As in~\cite{WHDG}, the energy loss probability $P(E_i \rightarrow E_f)$ is 
generalized to include both radiative and collisional energy loss and their 
fluctuations. However, a major difference between~\cite{WHDG} and the 
present study is that we here take into account both the 
radiative~\cite{MD_PRC} and collisional~\cite{MD_Coll} energy losses in 
a realistic {\it finite size dynamical} QCD medium.

To take into account geometric path length fluctuations in the energy loss 
probability, we use~\cite{WHDG}:
\begin{eqnarray}
P(&E_i& \rightarrow E_f = E_i-\Delta_{rad}-\Delta_{coll})=  \nonumber \\
&\;& \int dL \, P(L) \, P_{rad} (\Delta_{rad}; L) \otimes 
P_{coll}(\Delta_{coll}; L).
\label{fullconv}\end{eqnarray}
Here $P(L)$ is the distribution of path lengths traversed by hard scatterers 
in 0-5\% most central collisions, in which the lengths are weighted by the 
probability of production and averaged over azimuth. Note that currently two 
definitions for $P(L)$ (see~\cite{Simon_Thesis}), one from~\cite{WHDG} and the 
other from~\cite{Dainese}, are commonly used. Since these distributions are 
significantly different (see~\cite{Simon_Thesis}), we will use both of them 
in the analysis. Also, since $P(L)$ is a purely geometric quantity, it is the 
same for all jet varieties. 

$P_{rad} (\Delta_{rad}; L)$ and $P_{coll}(\Delta_{coll}; L)$ in Eq.~(\ref{fullconv})
are, respectively, the radiative and collisional energy loss probabilities. 
The procedure for including fluctuations of the radiative energy loss 
probability ($P_{rad} (\Delta_{rad}; L)$) due to gluon number fluctuations is 
discussed in detail in Ref.~\cite{Djordjevic:2005db,Djordjevic:2004nq}. Note 
that the procedure is here generalized to include the radiative energy loss in 
finite size dynamical QCD medium~\cite{MD_PRC,DH_PRL}, as well as a possibility 
for existence of finite magnetic mass~\cite{MD_MagnMass}; in particular, we 
extract the gluon radiation spectrum from Eq.~(10) in~\cite{MD_MagnMass}.
For collisional energy loss probability ($P_{coll}(\Delta_{coll}; L)$), the full 
fluctuation spectrum is approximated by a Gaussian centered at the average 
energy loss with variance $\sigma_{coll}^2 = 2 T \langle \Delta E^{coll}(p_\perp,L)\rangle $~\cite{Moore:2004tg,WHDG}. Here  $\Delta E^{coll}(p_\perp,L)$ is given by 
Eq.~(14) in~\cite{MD_Coll}, $T$ is the temperature of the medium, $p_\perp$ is 
the initial momentum of the jet, and $L$ is the length of the medium traversed 
by the jet. 

We note that, in the suppression calculations, we separately treat 
radiative from collisional energy loss; Consequently, we first calculate the 
modification of the quark and gluon spectrum due to radiative energy loss, 
and then due to collisional energy loss in QCD medium. This is a reasonable 
approximation when the radiative and collisional energy losses can be 
considered small (which is in the essence of the soft-gluon, soft-rescattering
approximation used in all energy loss calculations sofar~\cite{GLV,Gyulassy_Wang,Wiedemann,WW,DG_Ind,ASW,MD_TR,AMY,MD_PRC,DH_PRL}), and when collisional and 
radiative energy loss processes are decoupled from each other (which is the 
case in the HTL approach~\cite{RadVSColl} used in our energy loss 
calculations~\cite{MD_PRC,DH_PRL,MD_Coll}). Also, we assume that strong 
coupling constant $\alpha_S$ is fixed at $0.3$.

Finally, to obtain $\pi^0$ suppression from quark and gluon suppression, we 
use the following estimate~\cite{Vitev_pi,AG_pi}
\beqar
R_{AA}(\pi^0,p_\perp) \approx f_{g} R_{AA}(g,p_\perp) + (1 - f_{g}) R_{AA}(l,p_\perp) , 
\eeqar{RaaPions}
where $f_{g} \approx e^{- p_\perp /10.5 \, {\rm GeV}}$ is the fraction of pions with a 
given momentum $p_\perp$ that arise from gluon jet fragmentation, 
$R_{AA}(g,p_\perp)$ is the gluon suppression and $R_{AA}(l,p_\perp)$ is the light 
quark suppression.

To calculate D and B meson suppression, we will use both delta and 
Peterson~\cite{Peterson} fragmentation functions. Furthermore, to obtain 
single electron suppression, we will use the following estimate~\cite{DGVW}:
\beqar
R_{AA}(e^\pm,p_\perp) &\approx& \nonumber \\ 
&& \hspace{-2.5cm} 
\frac{R_{AA}(D,2 p_\perp) \frac{d\sigma_D(2 p_\perp)}{dp_\perp}  + 
R_{AA}(B,2 p_\perp) \frac{d\sigma_B (2 p_\perp)}{dp_\perp}}
{\frac{d\sigma_D (2 p_\perp)}{dp_\perp}+\frac{d\sigma_B (2 p_\perp)}{dp_\perp}} ,
\eeqar{RaaSingleE}
where $\frac{d\sigma_D(p_\perp)}{dp_\perp}$ ($\frac{d\sigma_B(p_\perp)}{dp_\perp}$)
is D (B) meson initial momentum distribution, and $R_{AA}(D,p_\perp)$ 
($R_{AA}(B,p_\perp)$) is the D (B) meson suppression. 
 
\section{Numerical results} 

In this section, we concentrate at 200 GeV Au+Au collisions at RHIC, and 
present our suppression predictions for light and heavy flavor observables. 
For this, we consider a quark-gluon plasma of temperature $T{\,=\,}225$\,MeV, 
with $N_f{\,=\,}2.5$ effective light quark flavors and strong interaction 
strength $\alpha_S{\,=\,}0.3$, as representative of average conditions 
encountered in Au+Au collisions at RHIC. For the light quarks we assume that 
their mass is dominated by the thermal mass $M{\,=\,}\mu/\sqrt{6}$, where 
$\mu{\,=\,}gT\sqrt{1{+}N_f/6}\approx 0.5$ GeV is the Debye screening mass. 
The gluon mass is taken to be $m_g=\mu/\sqrt{2}$. For the charm 
(bottom) mass we use $M{\,=\,}1.2$\,GeV ($M{\,=\,}4.75$\,GeV). 
%
\begin{figure*}
\epsfig{file=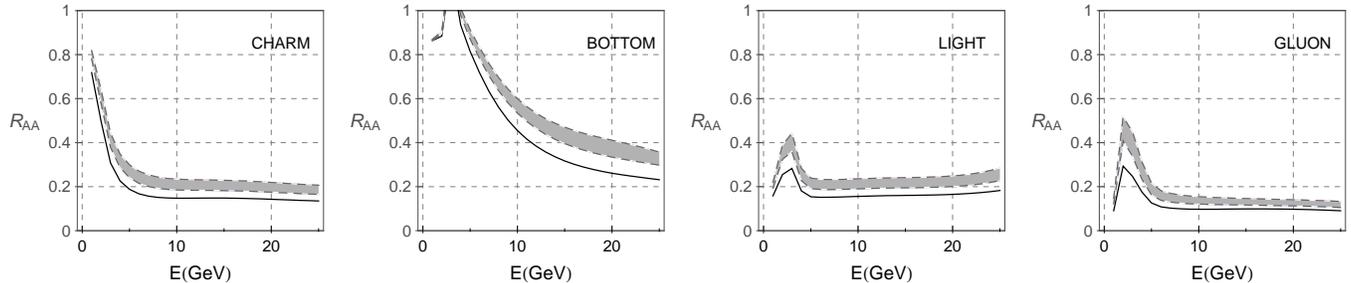,width=7.in,height=1.6in,clip=5,angle=0}
\vspace*{-0.5cm}
\caption{Quarks and gluon suppressions are presented as a function of initial 
jet energy for 200GeV Au+Au collisions at RHIC. The panels are obtained by 
using path length distributions from~\cite{WHDG}. On each panel full curves 
correspond to the case when magnetic mass is equal to zero. Gray bands 
correspond to the case when magnetic mass is non-zero (i.e.  $0.4 < \mu_M/\mu_E < 0.6$~\cite{Maezawa,Nakamura,Hart,Bak}), where the lower boundary corresponds 
to $\mu_M/\mu_E=0.4$ and the upper boundary corresponds to $\mu_M/\mu_E =0.6$.}
\label{QuarkRaa}
\end{figure*}
%
\begin{SCfigure*}
\epsfig{file=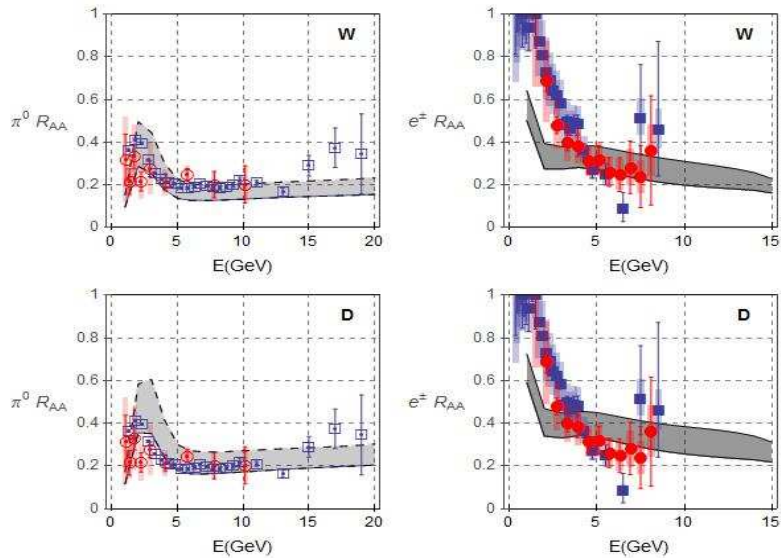,width=4.5in,height=3in,clip=5,angle=0}
\caption{Left panels show the comparison of pion suppression predictions with 
$\pi^0$ PHENIX~\cite{PHENIX_pi0} and STAR~\cite{STAR_pi0} experimental data 
from 200 GeV Au+Au collisions at RHIC. Right panels show the comparison of 
single electron suppression predictions with non-photonic single electron data
from PHENIX~\cite{PHENIX_e} and STAR~\cite{STAR_e} at 200 GeV Au+Au collisions.
For the two upper panels, and the two lower panels, the suppression 
predictions are obtained, respectively, by using path length distributions 
from~\cite{WHDG} (marked with ``W'') and~\cite{Dainese} (marked with ``D''). 
On each panel, the gray region corresponds to the case when $\mu_M \geq 0$ 
(i.e.  $0 < \mu_M/\mu_E < 0.6$), where the lower boundary corresponds 
to $\mu_M/\mu_E=0$ and the upper boundary corresponds to $\mu_M/\mu_E =0.6$.}
\label{PionsSingleE}
\end{SCfigure*}

Figure~\ref{QuarkRaa} shows momentum dependence of quarks and gluon 
suppressions at RHIC, obtained by using path length distributions 
from~\cite{WHDG}. We observe a clear hierarchy between the quarks and gluon 
suppressions: {\it i)} bottom quark is significantly less suppressed than 
charm quark; {\it ii)} charm and light quarks have similar suppressions for 
initial jet energies larger than 5 GeV; {\it iii)} gluons are significantly 
more suppressed than all types of quarks. This already observed/established 
hierarchy (see e.g.~\cite{Djordjevic:2004nq}) therefore remains valid for the 
case of dynamical QCD medium as well, despite the fact that both inclusion of 
path length fluctuations~\cite{WHDG} and dynamical effects~\cite{DH_PRL} into 
the suppression calculations tend to reduce the difference between different 
types of quarks and gluons. We also observe that inclusion of magnetic mass 
can decrease the jet suppression (for all types of quarks and gluons) from 
25-50$\%$, compared to the case of zero magnetic mass. While in 
Fig.~\ref{QuarkRaa} we show only the results by using path length distributions 
from~\cite{WHDG}, we observe somewhat lower suppression results when 
distribution from~\cite{Dainese} is used (data not shown); this is expected 
having in mind higher probability for lower path-lengths in~\cite{Dainese} 
compared to~\cite{WHDG} (see~\cite{Simon_Thesis}). 

To calculate D and B meson suppression, we used both delta and 
Peterson~\cite{Peterson} fragmentation functions, and observed that the 
choice of fragmentation function only marginally changes the value of 
suppression $R_{AA}$ (data not shown). The reason behind the small 
difference is that fragmentation functions do not significantly modify the 
distribution slopes~\cite{Djordjevic:2004nq}, due to which the meson 
suppression becomes insensitive to the choice of fragmentation function. For 
simplicity, we will therefore further use delta function fragmentation for 
the calculation of single electron suppression.

Figure~\ref{PionsSingleE} shows momentum dependence of pion and single 
electron suppressions at RHIC, obtained by using two path length distributions 
(from~\cite{WHDG} and from~\cite{Dainese}). The predictions are compared with 
the relevant PHENIX~\cite{PHENIX_pi0,PHENIX_e} and RHIC~\cite{STAR_pi0,STAR_e} 
experimental data at Au+Au collisions at RHIC. For both path length 
distributions we observe a reasonable agreement with the experimental data, 
which is moreover robust with respect to non-zero magnetic mass introduction.

Furthermore, from Fig.~\ref{PionsSingleE} we see that at jet energies above 
15 GeV, the predicted pion and single electron suppressions become very 
similar. This prediction is reasonable, since {\it i)} at high jet energies 
suppression patterns for all types of quarks become similar, and {\it ii)} 
above 10~GeV pion suppression is strongly dominated by light quark suppression 
(i.e. gluon contribution to pion suppression becomes 
negligible)~\cite{Vitev_pi,AG_pi}. This behavior is qualitatively different 
from the one below 10 GeV, where obtained suppressions are notably 
different (by $\lesssim 2$). It will be interesting to compare this predicted 
pattern with the upcoming high luminosity RHIC data.

\section{Conclusions} 

In this paper we calculated the suppression pattern of pions, D and B mesons 
and single electrons in central 200 GeV Au+ Au collisions at RHIC energies. 
The calculation is based on the radiative and collisional energy loss in a
finite size dynamical QCD medium, which is a key ingredient for obtaining 
reliable predictions for jet quenching in ultra-relativistic heavy ion 
collisions. This energy loss formalism was here integrated into a computational 
framework that includes multi-gluon and path length fluctuations. We obtained 
a reasonably good agreement between the generated suppression patterns and 
experimental data at Au+Au collisions at RHIC, and this agreement is robust 
with respect to introduction of finite magnetic mass. The agreement strongly 
suggests that the main deficiency responsible for the ``heavy flavor puzzle'' 
at RHIC was the {\it static} approximation, i.e. the fact that dynamical nature 
of plasma constituents was not taken into account. Predictions of the dynamical 
energy loss formalism remain to be tested against the upcoming high luminosity 
RHIC and LHC data.  

\medskip
{\em Acknowledgments:} 
This work is supported by Marie Curie International Reintegration Grant 
within the $7^{th}$ European Community Framework Programme 
(PIRG08-GA-2010-276913) and by the Ministry of Science and Technological 
Development of the Republic of Serbia, under projects No. ON171004 and 
ON173052.


\begin{references} 

\bibitem{Bjorken} J.D. Bjorken: FERMILAB-PUB-82-059-THY (1982).

\bibitem{Brambilla} N. Brambilla et al., Preprint hep-ph/0412158 (2004).

\bibitem{Gyulassy} M. Gyulassy, Lect. Notes Phys. {\bf 583}, 37 (2002).

\bibitem{DBLecture} D. d'Enterria, B. Betz, Lect. Notes Phys. {\bf 785}, 285 
(2010). 

\bibitem{DGVW} M. Djordjevic, M. Gyulassy, S. Wicks and R. Vogt, Phys. Lett. 
B {\bf 632}, 81 (2006).

\bibitem{RGH} R. Rapp, V. Greco and H. van Hees, Nucl. Phys. A {\bf 774}, 
685 (2006).

\bibitem{ADSW} N. Armesto, A. Dainese, C. A. Salgado and U. A. Wiedemann, 
Phys. Rev. D {\bf 71}, 054027 (2005).

\bibitem{PHENIX_pi0} A. Adare {\it et al.} [PHENIX collaboration], 
Phys. Rev. Lett. {\bf 101}, 232301 (2008).

\bibitem{PHENIX_e}  A. Adare {\it et al.} [PHENIX collaboration], 
Phys. Rev. Lett. {\bf 98}, 172301 (2007).

\bibitem{STAR_pi0} B.I. Abelev {\it et al.} [STAR collaboration], 
Phys. Rev. C {\bf 80} 44905 (2009).

\bibitem{STAR_e} B.I. Abelev {\it et al.} [STAR collaboration], 
Phys. Rev. Lett. {\bf 98}, 192301
(2007).

\bibitem{Gyulassy_viewpoint} M. Gyulassy, Physics {\bf 2}, 107 (2009).

\bibitem{SE_puzzle} M. Djordjevic, J. Phys. G {\bf 32}, S333 (2006).

\bibitem{suppression} M. Guylassy, I. Vitev, X. N. Wang and B. W. Zhang, in 
Quark Gluon Plasma 3, edited by R. C. Hwa and X. N. Wang, p. 123 
(World Scientific, Singapore, 2003)  

\bibitem{BDMS} 
  R.~Baier, Yu.~L.~Dokshitzer, A.~J.~Mueller and D.~Schiff, 
  Phys. Rev. C {\bf 58}, 1706 (1998).

\bibitem{BSZ} 
  R.~Baier, D.~Schiff, B.~G.~Zakharov, 
  Ann. Rev. Nucl. Part. Sci. {\bf 50}, 37 (2000).
 
\bibitem{KW:2004} 
  A.~Kovner and U.~A.~Wiedemann, in
  {\it Quark Gluon Plasma 3}, edited by R.C. Hwa and X.N. Wang,
  p.~192 (World Scientific, Singapore, 2003).

\bibitem{MD_PRC}  M. Djordjevic, Phys. Rev. C {\bf 80}, 064909 (2009).

\bibitem{DH_PRL}  M. Djordjevic and U. Heinz, Phys. Rev. Lett. {\bf 101}, 
022302 (2008). 

\bibitem{DG_Ind}
  M.~Djordjevic and M.~Gyulassy, Phys.\ Lett.\ B \ {\bf 560}, 37 (2003); 
  and Nucl.\ Phys.\ A {\bf 733}, 265 (2004).

\bibitem{GLV} 
  M.~Gyulassy, P.~Levai and I.~Vitev,
  Nucl.\ Phys.\ B {\bf 594}, 371 (2001).

\bibitem{Gyulassy_Wang} 
  M.~Gyulassy and X.~N.~Wang, 
  Nucl.\ Phys.\ B {\bf 420}, 583 (1994);
  X.~N.~Wang, M.~Gyulassy and M.~Plumer, Phys. Rev. D {\bf 51} (1995)  3436. 

\bibitem{Wiedemann} 
  U.A.~Wiedemann, Nucl. Phys. B {\bf 588}, 303 (2000); 
  and Nucl. Phys. B {\bf 582}, 409 (2000). 

\bibitem{WW} 
  E. Wang and X. N. Wang, Phys. Rev. Lett. {\bf 87} 142301, (2001). 
  X. N. Wang and X. F. Guo, Nucl. Phys. A {\bf 696}, 788 (2001); 
  X. F. Guo and X. N. Wang, Phys. Rev. Lett. {\bf 85} 3591, (2000). 

\bibitem{ASW} 
  N. Armesto, C.~A.~Salgado and U. A. Wiedemann,  
  Phys. Rev. D {\bf 69}, 114003 (2004). 

\bibitem{MD_TR}
M. Djordjevic, Phys. Rev. C {\bf 73}, 044912 (2006). 

\bibitem{MD_MagnMass} M. Djordjevic, arXiv:1105.4359 [nucl-th] 

\bibitem{Maezawa} Yu. Maezawa et al. [WHOT-QCD Collaboration], 
Phys. Rev. D {\bf 81} 091501 (2010); 
Yu. Maezawa et al. [WHOT-QCD Collaboration], PoS Lattice 
194 (2008).

\bibitem{Nakamura} A. Nakamura, T. Saito and S. Sakai, Phys. Rev. D {\bf 69},
014506 (2004).

\bibitem{Hart} A. Hart, M. Laine and O. Philipsen, Nucl. Phys. B {\bf 586},
443 (2000).

\bibitem{Bak} D. Bak, A. Karch and L. G. Yaffe, JHEP {\bf 0708}, 049
(2007).

\bibitem{MD_Coll}  M. Djordjevic, Phys. Rev. C {\bf 74}, 064907 (2006). 

\bibitem{GLV_suppress}
M. Gyulassy, P. Levai and I. Vitev,
Phys.\ Lett.\ B {\bf 538}, 282 (2002).

\bibitem{WHDG}  S. Wicks, W. Horowitz, M. Djordjevic and M. Gyulassy, Nucl.
Phys. A {\bf 784}, 426 (2007). 

\bibitem{Djordjevic:2005db}
  M.~Djordjevic, M.~Gyulassy, R.~Vogt and S.~Wicks,
  Phys. Lett. B {\bf 632} 81 (2006). 

\bibitem{Cacciari:2005rk}
  M.~Cacciari, P.~Nason and R.~Vogt,
  Phys. Rev. Lett. {\bf 95} 122001 (2005). 

\bibitem{MNR}
M. L. Mangano, P. Nason and G.Ridolfi,
Nucl.\ Phys.\ B {\bf 373}, 295 (1992).

\bibitem{Vitev:2002pf}
I.~Vitev and M.~Gyulassy,
Phys.\ Rev.\ Lett.\  {\bf 89}, 252301 (2002).
 
\bibitem{Vogt}
R. Vogt, 
Int.\ J.\ Mod.\ Phys.\ E {\bf 12}, 211 (2003).

\bibitem{Lai:1999wy}
  H.~L.~Lai {\it et al.}  [CTEQ Collaboration],
  Eur.\ Phys.\ J.\ C {\bf 12}, 375 (2000).

\bibitem{Simon_Thesis} S. Wicks, PhD thesis, AAT-3333486, 
PROQUEST-1614268891, 2008. 338pp

\bibitem{Dainese} A. Dainese, Eur. Phys. J. C {\bf 33} 495 (2004).

\bibitem{Djordjevic:2004nq}
  M.~Djordjevic, M.~Gyulassy and S.~Wicks,
  Phys.\ Rev.\ Lett.\  {\bf 94}, 112301 (2005).

\bibitem{Moore:2004tg}
  G.~D.~Moore, D.~Teaney,
  Phys.\ Rev.\ C {\bf 71}, 064904 (2005).

\bibitem{AMY}
P. Arnold, G. D. Moore and L. G. Yaffe, JHEP {\bf 0111}, 057 (2001); 
JHEP {\bf 0206}, 030 (2002); JHEP {\bf 0301}, 030 (2003).

\bibitem{RadVSColl} M. Djordjevic, Nucl. Phys. A {\bf 783} 197 (2007). 

\bibitem{Vitev_pi} I. Vitev, Phys. Lett. B {\bf 606}, 303 (2005). 

\bibitem{AG_pi} A. Adil and M. Gyulassy, Phys. Lett. B {\bf 602} (2004) 52.

\bibitem{Peterson}
C. Peterson, D. Schlatter, I. Schmitt and P. M. Zerwas
Phys.\ Rev.\ D {\bf 27}, 105 (1983). 

\end{references}
\end{document}